\documentclass[prl,twocolumn,floatfix,showpacs ]{revtex4-1}
\UseRawInputEncoding
\usepackage{amsmath}
\usepackage{amssymb}
\usepackage{graphicx}
\usepackage{dcolumn}
\usepackage{bm}
\usepackage[colorlinks=true,linkcolor=blue,anchorcolor=blue, citecolor=cyan,urlcolor=cyan]{hyperref}
\usepackage[mathlines]{lineno}
\usepackage{ulem}
\usepackage{epstopdf}

\begin{document}
\title{Achieving fully-compensated ferrimagnetism through  two-dimensional heterojunctions}
\author{San-Dong Guo$^{1}$$^{\textcolor[rgb]{0.00,0.00,1.00}{\dagger}}$}
\email{sandongyuwang@163.com}
\author{Junjie He$^{2}$}
\thanks{These authors contributed equally to this work.}
\author{Yee Sin Ang$^{3}$}
\affiliation{$^{1}$School of Electronic Engineering, Xi'an University of Posts and Telecommunications, Xi'an 710121, China}
\affiliation{$^{2}$KFMCH, Faculty of Science, Charles University, Prague 12843, Czech Republic}
\affiliation{$^3$Science, Mathematics and Technology (SMT), Singapore University of Technology and Design (SUTD), 8 Somapah Road, Singapore 487372, Singapore}
\begin{abstract}
In addition to altermagnets, fully-compensated ferrimagnets are another category of collinear magnetic materials that possess zero-net total magnetic moment and exhibit spin-splitting,  making them promising for low-energy spintronics, high-density data storage and high-sensitivity sensors. Although many methods, such as alloying, external electric field, Janus engineering, ferroelectric field and spin ordering, have been proposed to achieve fully-compensated ferrimagnetism, these approaches either face experimental difficulties or  produce a small spin-splitting   or are volatile. Here, we propose to  form vertical heterostructures by stacking two different but equally magnetized two-dimensional ferromagnetic materials. If an A-type antiferromagnetic  ordering is satisfied, a fully compensated ferrimagnet can be formed. This vertical heterostructure approach is insensitive to lattice matching and stacking manner, thus being more conducive to experimental realization. Through first-principles calculations, we verify our proposal with several examples, focusing in particular on $\mathrm{CrI_3}$/$\mathrm{CrGeTe_3}$ heterojunction composed of experimentally synthesized $\mathrm{CrI_3}$ and $\mathrm{CrGeTe_3}$ monolayers. The calculations show that $\mathrm{CrI_3}$/$\mathrm{CrGeTe_3}$ is a fully-compensated ferrimagnet, with pronounced spin-splitting, and that tensile strain is more favorable for achieving fully-compensated ferrimagnetism. Our work provides an experimentally feasible strategy for realizing fully-compensated ferrimagnetism, thereby further advancing the development of this field.

\end{abstract}
\maketitle
\textcolor[rgb]{0.00,0.00,1.00}{\textbf{Introduction.---}}
Zero-net-magnetization systems are emerging, which offer superior spintronic performance with  ultrahigh data densities, immunity to external perturbations, and femtosecond-scale writing speeds\cite{k1,k2}. Although conventional antiferromagnets possess zero net magnetization, the absence of spin-splitting severely limits their practical applications. Recently, alternagnets  have garnered significant attentions in the field of magnetism, because they  not only possess the zero-net-magnetization of antiferromagnets in real space, but also inherit the spin-splitting characteristics of ferromagnets in momentum space\cite{k4,k5,k6,k7,k8,k9,k10,zg1,qq3}.
Alternagnets  provide an ideal platform for both fundamental science and the practical applications of next-generation information technologies.

In addition to altermagnets, fully-compensated ferrimagnets constitute another category of collinear magnetic materials, characterized by a zero-net magnetic moment and the presence of spin-splitting\cite{f1,f2,f3,f5,f6,f7,f8,f9,f4,zg2}.  The two spin sublattices of   conventional antiferromagnets or  alternagnets are connected by either space inversion/translation  ($P/\tau$) or rotational/mirror ($C/M$) symmetry,  but the two spin sublattices  in fully-compensated ferrimagnets are not connected by any symmetry. Fully-compensated ferrimagnets, like alternagnets, can also exhibit a range of phenomena, including the anomalous Hall and Nernst effects, non-relativistic spin-polarized currents, and the magneto-optical Kerr effect\cite{f4}.

\begin{figure}[t]
    \centering
    \includegraphics[width=0.45\textwidth]{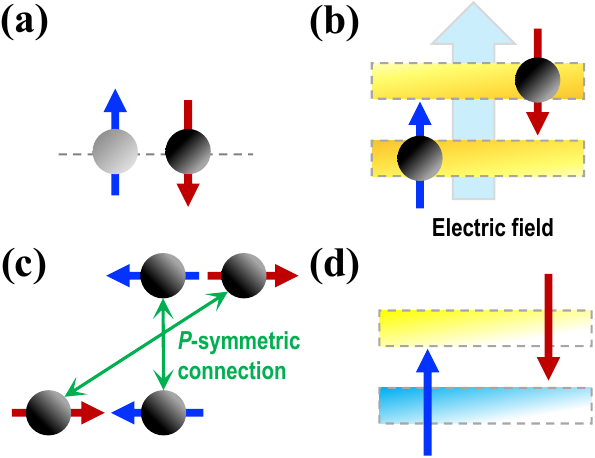}
    \caption{(Color online) (a): the isovalent alloying can induce fully-compensated ferrimagnetism in $PT$-antiferromagnets or altermagnets. The black  and gray spheres  represent, for example, 4$d$ and 3$d$ elements from the same chemical group, respectively. (b): an  electric field   can induce fully-compensated ferrimagnetism  in $PT$-antiferromagnets or altermagnets with the layer-dependent opposite spin polarization, which can be an external electric field, the built-in field from a Janus structure, or the internal field of ferroelectric polarization. (c): the spin ordering-induced fully-compensated ferrimagnetism with the system possessing lattice $P$ symmetry. (d): by stacking two different monolayers with the same total magnetic moments together to form the so-called heterojunction, the fully-compensated ferrimagnetism can be induced. In (a), (b), (c), and (d), blue arrows represent spin up, while red arrows represent spin down. In (b), the big arrow represents the electric field. In (c), the green arrows represent the $P$-symmetric connection between two magnetic atoms. }\label{a}
    \end{figure}

Fully-compensated ferrimagnets can be realized through alloying (see \autoref{a} (a)), mainly focusing on bulk materials\cite{f1,f2,f3}.
For example, two magnetic atoms with opposite spin polarizations are from 4$d$ and 3$d$ elements from the same chemical group, respectively.
If a two-dimensional (2D) system has spin-layer coupling (A-type antiferromagnetic (AFM) ordering), an electric field can induce fully-compensated ferrimagnetism  (see \autoref{a} (b)), which can be an external electric field, the built-in field from a Janus structure, or the internal field of ferroelectric polarization\cite{f4,zg2, gsd1,gsd2,gsd3,gsd4,qq5,qq6}.  Very recently, fully-compensated ferrimagnetism has been experimentally achieved in bilayer  $\mathrm{CrPS_4}$ by a perpendicular external electric field\cite{nn}.  Fully-compensated ferrimagnetism can also be realized by engineering the spin ordering rather than modifying the lattice structure (see \autoref{a} (c))\cite{qq7}. Although many strategies for achieving fully-compensated ferrimagnetism have been proposed, some face experimental challenges (such as that induced by Janus engineering\cite{f4,zg2, gsd1,gsd2}), some induce relatively small spin-splitting (such as that induced by sliding ferroelectrics\cite{gsd4,qq5,qq6}), and some are volatile  (such as that induced by external electric field\cite{f4,zg2, gsd1,gsd2,gsd3}). Here, we propose to achieve fully compensated ferrimagnetism by stacking two different ferromagnetic (FM) monolayers with the same total magnetic moment to form a heterojunction. Our proposal not only can produce a large non-volatile spin-splitting but also may be relatively easy to implement experimentally.

\textcolor[rgb]{0.00,0.00,1.00}{\textbf{Approach.---}}
In a fully-compensated ferrimagnet, the two sublattices with opposite spin polarizations are not connected by any symmetry\cite{f4}. That is to say, the surrounding environments of the two magnetic atoms with opposite spin polarizations are different. In fact, the magnetic moment of a magnetic atom can be replaced by the total magnetic moment of a primitive cell. If we periodically arrange two primitive cells made of different materials but with the same total magnetic moment, and require that the total magnetic moments of these two primitive cells are opposite, then the condition for forming a fully-compensated ferrimagnet as mentioned earlier can be satisfied. The materials involved should preferably be FM, as this is more conducive to the formation of a fully-compensated ferrimagnet with a large global spin-splitting. Of course, the materials involved can also be conventional antiferromagnets and  altermagnets, but this may not be favorable for large global spin-splitting.

The 2D materials should be the optimal candidate materials, and periodic arrangement in the form of heterostructure is also the simplest or most experimentally feasible method.  As shown in \autoref{a} (d), stacking two different monolayers with the same total magnetic moment in an A-type AFM arrangement, which is also known as a heterostructure approach, typically breaks the  [$C_2$$\parallel$$O$]  (The $C_2$ is the two-fold rotation perpendicular to the spin axis in spin space, and $O$  means $P$/$C$/$M$  in lattice space) symmetry and achieves a fully-compensated ferrimagnet. If the two monolayers are altermagnets, such stacking may result in the bilayer still being a altermagnet. To experimentally verify our proposal, we conduct a detailed discussion on $\mathrm{CrI_3}$/$\mathrm{CrGeTe_3}$ heterostructure using experimentally synthesized monolayer $\mathrm{CrI_3}$ and $\mathrm{CrGeTe_3}$\cite{p2,p3} as building blocks, and finally provide a brief illustration with some other examples.

 \begin{figure}[t]
    \centering
    \includegraphics[width=0.45\textwidth]{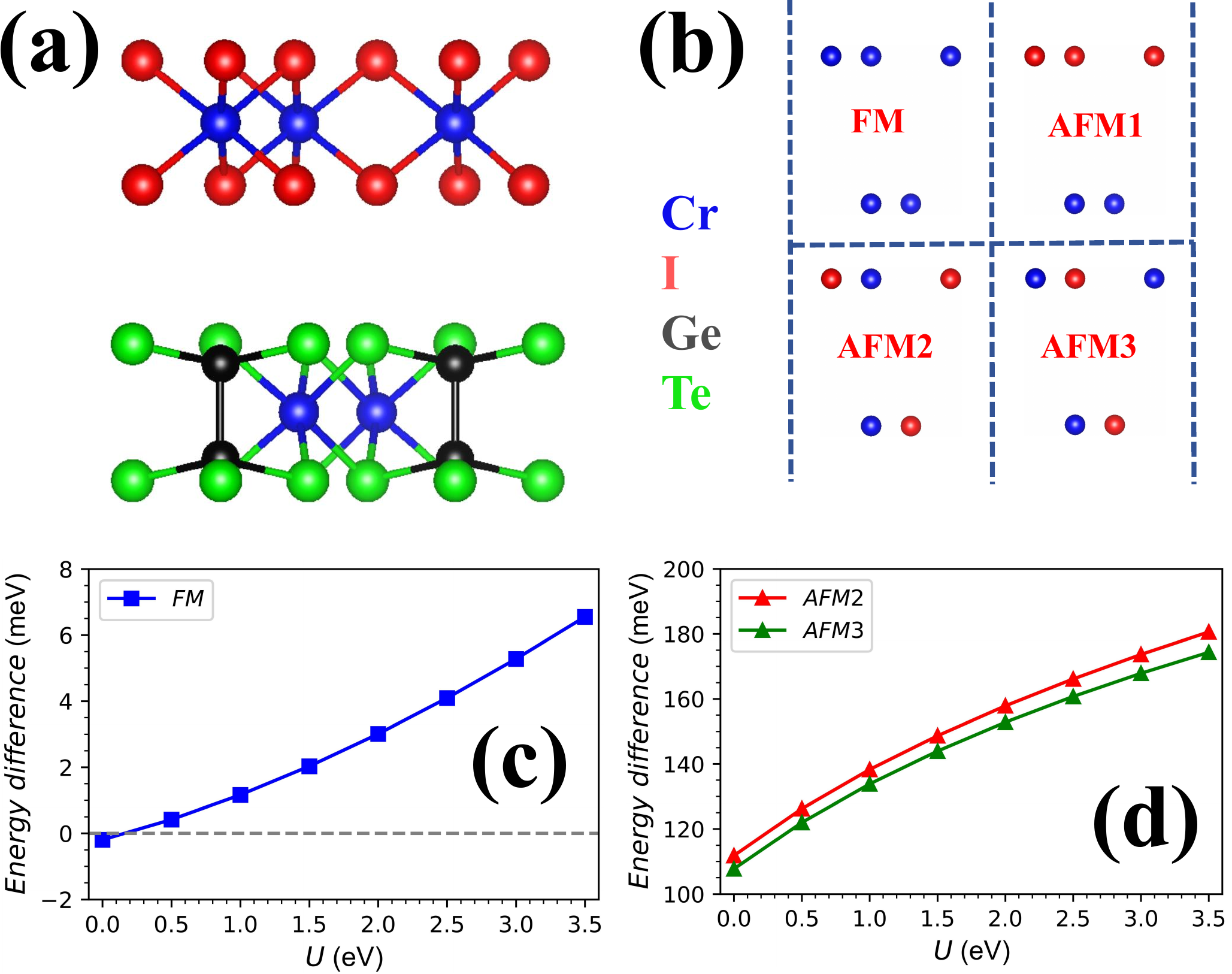}
    \caption{(Color online)For AB-stacked $\mathrm{CrI_3}$/$\mathrm{CrGeTe_3}$ heterojunction, (a): the lattice structure; (b): four different magnetic configurations: FM, AFM1, AFM2 and AFM3; (c): the energy of the FM configuration as a function of $U$ with the AFM1 configuration as the reference; (d): the energies of the AFM2 and AFM3 configurations as a function of  $U$ with the AFM1 configuration as the reference. In (a), the blue, red, black, and green spheres represent Cr, I, Ge, and Te atoms, respectively. In (b), the blue and red small spheres represent Cr atoms with spin up and spin down, respectively.}\label{b}
\end{figure}

\begin{figure*}[t]
    \centering
    \includegraphics[width=0.85\textwidth]{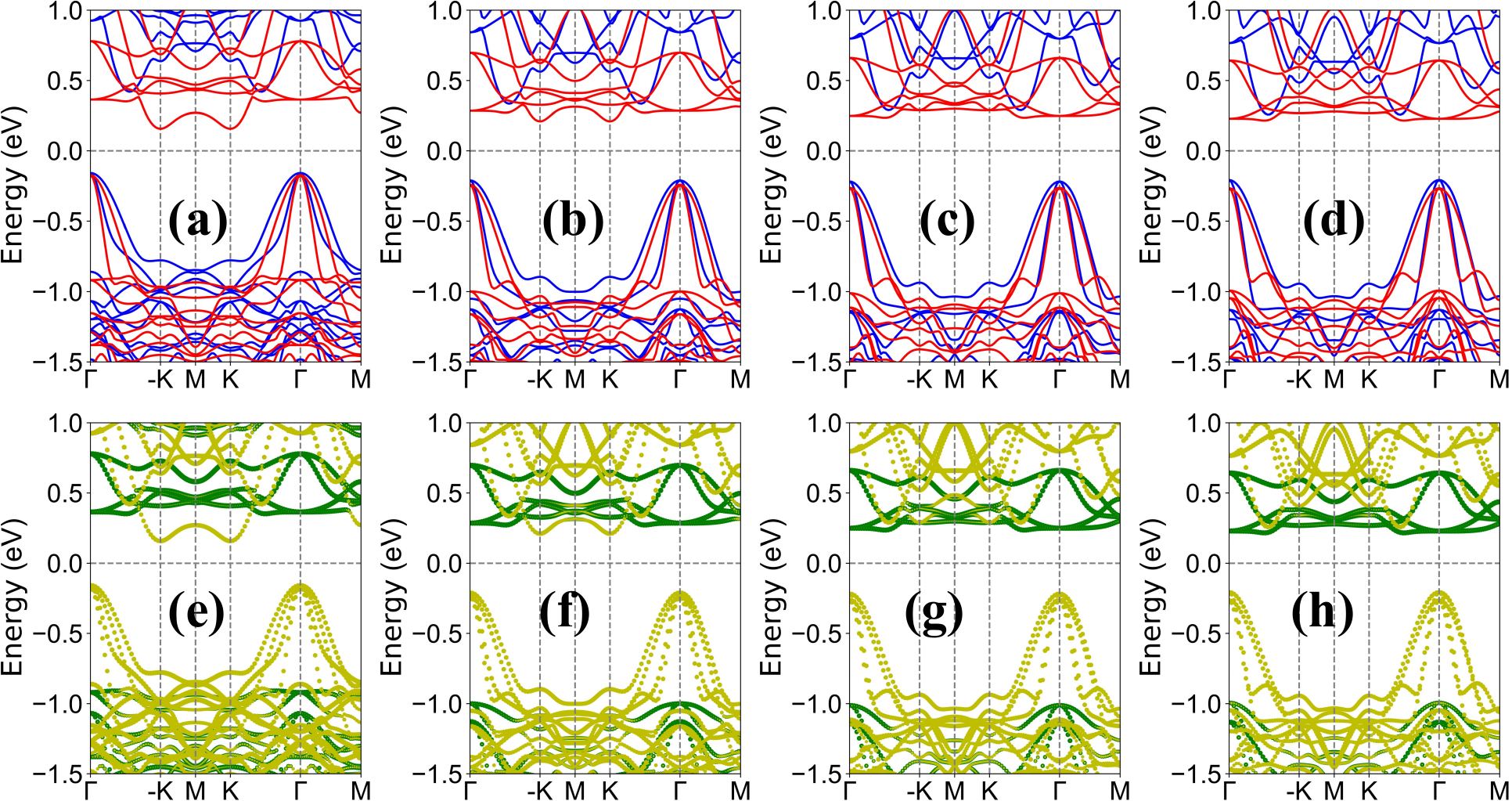}
    \caption{(Color online)For AB-stacked $\mathrm{CrI_3}$/$\mathrm{CrGeTe_3}$ heterojunction, the spin-polarized band structures (a, b, c, d) and the layer-projected band structures  (e, f, g, h) at $U$ = 0.00 (a, e), 1.00 (b, f), 2.00 (c, g), 3.00 (d, h) eV.  In (a, b, c, d),  the spin-up
and spin-down channels are depicted in blue and red. In (e, f, g, h),
      the yellow and green represent  $\mathrm{CrGeTe_3}$- and  $\mathrm{CrI_3}$-layer characters.}\label{c}
\end{figure*}

\textcolor[rgb]{0.00,0.00,1.00}{\textbf{Computational detail.---}}
We perform the spin-polarized first-principles calculations  within density functional theory (DFT)\cite{1} and  the projector augmented-wave (PAW) method by using the Vienna Ab Initio Simulation Package (VASP)\cite{pv1,pv2,pv3}.  The  Perdew-Burke-Ernzerhof generalized gradient approximation (GGA)\cite{pbe} is used  as the exchange-correlation functional. We add Hubbard correction $U$ for $d$-orbitals of both Cr and Y atoms within the
rotationally invariant approach proposed by Dudarev et al\cite{du}. The dispersion-corrected DFT-D3 method\cite{dft3} is adopted to describe the van der Waals (vdW) interactions. The calculations are carried out with the kinetic energy cutoff  of 500 eV,  total energy  convergence criterion of  $10^{-8}$ eV, and  force convergence criterion of 0.01 $\mathrm{eV{\AA}^{-1}}$.  A vacuum layer exceeding 20 $\mathrm{{\AA}}$ along the $z$-direction is employed to eliminate spurious interactions between periodic images. The Brillouin zone (BZ) is sampled with a 12$\times$12$\times$1 (21$\times$21$\times$1) Monkhorst-Pack $k$-point meshes for both structural relaxation and electronic structure calculations of  $\mathrm{CrI_3}$/$\mathrm{CrGeTe_3}$ ($\mathrm{YBr_2}$/$\mathrm{YCl_2}$) heterojunction.

\textcolor[rgb]{0.00,0.00,1.00}{\textbf{Material realization.---}}
Monolayer $\mathrm{CrI_3}$ has  been synthesized experimentally\cite{p2}, which is composed of three monatomic planes in the sequence I-Cr-I and crystallizes in the  $P\bar{3}1m$ space group (No.162).   The FM state of   $\mathrm{CrI_3}$ is the most stable magnetic
configuration with  total magnetic moments of  6 $\mu_B$ per unit cell,  and   the  optimized  equilibrium lattice constants are  $a$=$b$=7.00 $\mathrm{{\AA}}$ within GGA.
Monolayer $\mathrm{CrGeTe_3}$  can be obtained by replacing I in $\mathrm{CrI_3}$ with Te and additionally adding two Ge atoms at the center of the voids.
 Each of the two Ge atoms in the unit cell is only threefold coordinated to the Te atoms. These modifications result in $\mathrm{CrGeTe_3}$ and $\mathrm{CrI_3}$  having the same crystallographic space group. The $\mathrm{CrGeTe_3}$ possesses also the FM ground state\cite{p3}  with  total magnetic moments of  6 $\mu_B$, and the   optimized  equilibrium lattice constants are  $a$=$b$=6.91 $\mathrm{{\AA}}$ within GGA. These two monolayers are very suitable for forming heterojunction to validate our proposal.

We stack $\mathrm{CrI_3}$ and $\mathrm{CrGeTe_3}$ to form heterostructures, considering both AA (FIG.S1\cite{bc}) and AB (\autoref{b} (a)) stacking configurations. They crystallize  in the  $P31m$  (No.157) and  $P3$  (No.143) space group, respectively.  The  optimized  equilibrium lattice constants are  $a$=$b$=6.87 $\mathrm{{\AA}}$ within GGA for  both AA  and AB stacking configurations.
To determine the magnetic ground state, we have examined four magnetic configurations, namely FM, AFM1, AFM2, and AFM3, which are shown in \autoref{b} (b).
Because the energies of AFM2 and AFM3 are much higher than those of FM and AFM1, we only present here the energy comparisons of FM and AFM1 for AA and AB stacking.   The calculations reveal that the AB stacking possesses lower energy, and the AFM1 configuration is the one required as previously proposed. Therefore, taking the AFM1 configuration of the AB stacking as the reference, the energies of the FM and AFM1 configurations for AA stacking, and the energy of the FM configuration for AB stacking are 26.39, 21.93, and -0.28 meV, respectively. These results show that the FM ordering of the AB stacking has the lowest energy.
Because the energy of AA stacking is significantly higher than that of AB stacking, we focus on the AB stacking case in the following discussions.

Nevertheless, electron correlation has a significant impact on the magnetic configuration, electronic structure, and topological properties of 2D  magnetic materials\cite{p4}. Therefore, we investigate the influence of electron correlation $U$ on the electronic properties of  AB-stacked $\mathrm{CrI_3}$/$\mathrm{CrGeTe_3}$ heterojunction. With AFM1 as the reference, the energies of FM, AFM2, and AFM3 as functions of $U$ are plotted in \autoref{b} (c) and (d).
Within the range of $U$ considered, the energies of AFM2 and AFM3 are much higher than those of FM and AFM1. When $U$ is just greater than 0.2 eV, AFM1 will become the ground state, which is precisely the magnetic configuration we proposed earlier to achieve fully-compensated ferrimagnetism. It is also clearly seen that increasing the strength of electron correlation is conducive to stabilizing the AFM1 ordering. Whin AFM1 ordeing, the energy band structures with spin- and  layer-characteristic projection  at  representative $U$=0.00,  1.00, 2.00 and 3.00  eV  are plotted in \autoref{c} (For comparison, the band structure for the case of $U$=0.00 eV is also plotted with the AFM1 ordering.).
\begin{figure*}[t]
    \centering
    \includegraphics[width=0.85\textwidth]{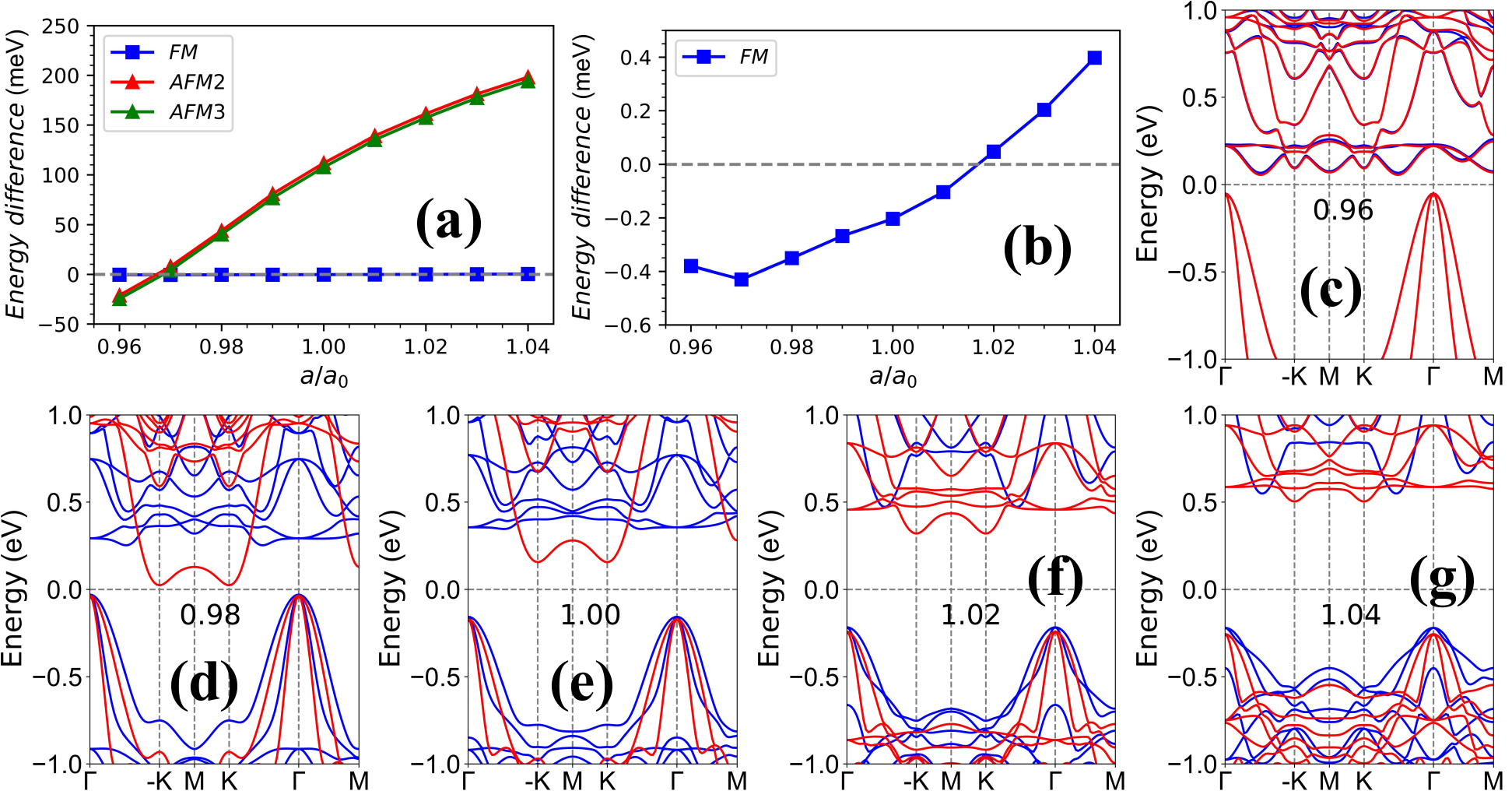}
    \caption{(Color online)For AB-stacked $\mathrm{CrI_3}$/$\mathrm{CrGeTe_3}$ heterojunction by GGA ($U$=0.00 eV),  (a): the energies of the FM, AFM2, AFM3 configurations as a function of $a/a_0$ with the AFM1 configuration as the reference;  (b): the energy of the FM configuration as a function of  $a/a_0$ with the AFM1 configuration as the reference. (c, d, e, f, g): the spin-polarized band structures at $a/a_0$=0.96, 0.98, 1.00, 1.02 and 1.04 with AFM3, FM, FM, AFM1 and AFM1 orderings, and  the spin-up and spin-down channels are depicted in blue and red. }\label{d}
\end{figure*}

Within the range of $U$ considered, all the total magnetic moments of  $\mathrm{CrI_3}$/$\mathrm{CrGeTe_3}$ heterojunction  are strictly 0.00 $\mu_B$, which can be guaranteed by the existing band gap\cite{f4}.
The total magnetic moment can also be calculated by subtracting the number of occupied electrons with spin down from the number of occupied electrons with spin up, which can be confirmed by the integrated densities of states (IDOSs).  For the two spin channels of fully-compensated ferrimagnets, IDOSs are not the same across the entire energy range, but it must be the same within the band gap\cite{f4}.  According to FIG.S2\cite{bc}, at the representative value of $U$=1.00 eV, the values of IDOSs for the two spin channels are indeed the same within the band gap, further ensuring a zero-net total magnetic moment. For fully-compensated ferrimagnets, the absolute values of the magnetic moments of the magnetic atoms are not strictly equal, which distinguishes them from  $PT$-antiferromagnets and altermagnets\cite{zg2}.  For example $U$=1.00 eV, the absolute values of the magnetic moments of the two Cr atoms in  $\mathrm{CrGeTe_3}$ layer are both approximately 3.16 $\mu_B$, while those of the two Cr atoms in  $\mathrm{CrI_3}$ layer are both approximately 3.10 $\mu_B$. This is because the two types of Cr atoms are not symmetrically connected.

In all cases of $U$, there is a pronounced global spin-splitting in $\mathrm{CrI_3}$/$\mathrm{CrGeTe_3}$ heterojunction. For altermagnets, there is no spin-splitting at the $\Gamma$ point, while fully-compensated ferrimagnets can exhibit spin splitting\cite{k4,f4}. According to \autoref{c}, there is a pronounced spin-splitting at the $\Gamma$ point, which is consistent with the requirements of a fully-compensated ferrimagnet.  The zero-net total magnetic moment and spin-splitting ensure that $\mathrm{CrI_3}$/$\mathrm{CrGeTe_3}$ heterojunction is indeed a fully-compensated ferrimagnet when $U$ is greater than 0.2 eV.
When $U$ is relatively small, the conduction band and valence band of $\mathrm{CrGeTe_3}$ both lie within the band gap of $\mathrm{CrI_3}$, forming  Type-I heterojunction (see \autoref{c} (e, f)). When $U$ increases, both the valence and conduction bands of $\mathrm{CrGeTe_3}$ are positioned above the valence and conduction bands of $\mathrm{CrI_3}$, resulting in a Type-II heterojunction (see \autoref{c} (g, h)).

Even when $U$ is less than 0.2 eV in $\mathrm{CrI_3}$/$\mathrm{CrGeTe_3}$ heterojunction, the fully-compensated ferrimagnetism can be achieved through strain engineering. We use $a/a_0$ (The $a_0$ and $a$ represent the lattice parameters without strain and with applied strain, respectively.) to simulate strain, where $a/a_0$$<$1 represents compressive strain, and $a/a_0$$>$1 represents tensile strain.
With AFM1 as the reference, the energies of FM, AFM2, and AFM3 as functions of $a/a_0$ are shown in \autoref{d} by GGA ($U$=0.00 eV).
Within the range of strain considered, when $a/a_0$ is less than 0.968, AFM3 is the ground state; when $a/a_0$ is greater than 0.968 but less than 1.016, FM has the lowest energy; and when $a/a_0$ is greater than 1.016, AFM1 is in the ground state.
Whin the magnetic  ordeing of  the ground state, the energy band structures with spin-characteristic projection  at  representative $a/a_0$=0.96,  0.98, 1.00, 1.02 and 1.04  are plotted in \autoref{d}.

When $a/a_0$ is greater than 1.106, $\mathrm{CrI_3}$/$\mathrm{CrGeTe_3}$ heterojunction exhibits a pronounced spin-splitting, and its total magnetic moment is 0.00 $\mu_B$, thus becoming a fully-compensated ferrimagnet even if $U$=0.00 eV.  When $\mathrm{CrI_3}$/$\mathrm{CrGeTe_3}$ heterojunction is in the FM state, it remains a semiconductor with a total magnetic moment of 12 $\mu_B$, which is precisely the sum of the total magnetic moments of the two independent monolayers. As we proposed earlier, even  stacking two different antiferromagnets can also achieve fully-compensated ferrimagnetism. When $a/a_0$ is less than 0.968, $\mathrm{CrI_3}$/$\mathrm{CrGeTe_3}$ heterojunction is in such a situation, characterized by a zero-net total magnetic moment and weak spin-splitting (The expanded conduction band near the Fermi level is shown in FIG.S3.\cite{bc}).  This example further illustrates that stacking two different AFM monolayers to achieve fully-compensated ferrimagnetism is not conducive to a large spin-splitting.
\begin{figure}[t]
    \centering
    \includegraphics[width=0.45\textwidth]{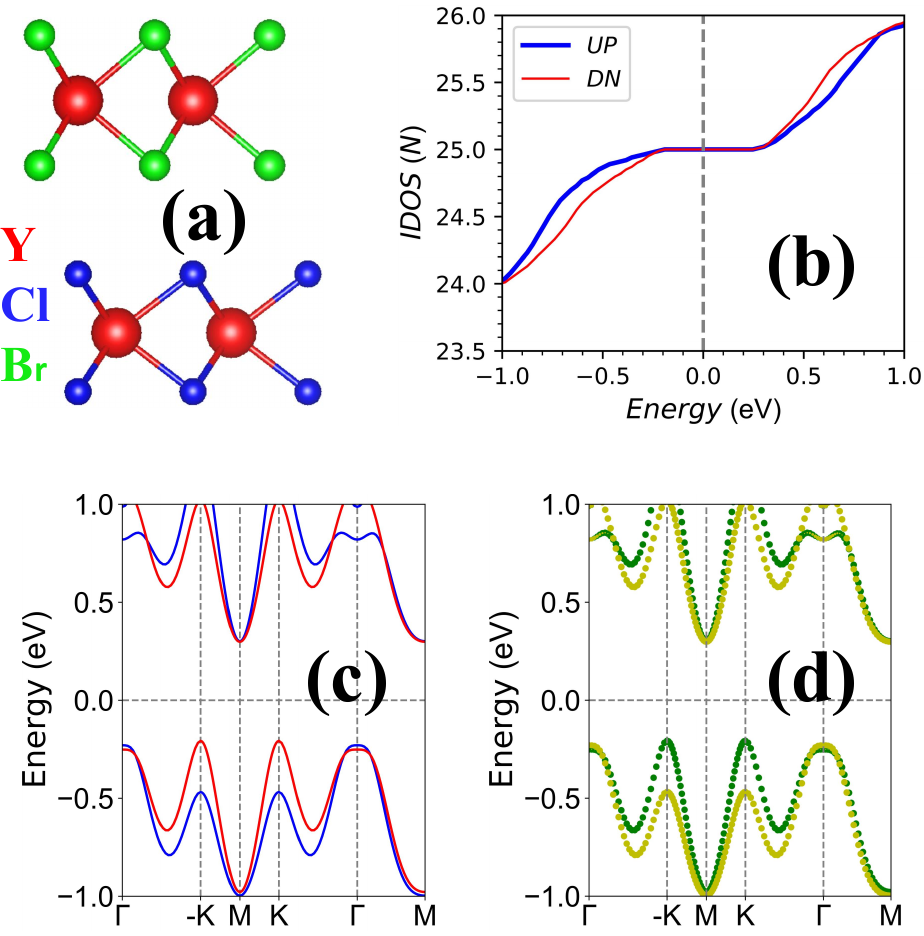}
    \caption{(Color online)For AB-stacked $\mathrm{YBr_2}$/$\mathrm{YCl_2}$ heterojunction, (a): the crystal structures;  (b): the integrated densities of states;  (c):  the spin-polarized band structures; (d):  the layer-projected band structures.   In (c),  the spin-up
and spin-down channels are depicted in blue and red. In (d),
      the yellow and green represent  $\mathrm{YCl_2}$- and  $\mathrm{YBr_2}$-layer characters. }\label{f}
\end{figure}

We have also considered other heterostructures to illustrate the universality of our proposal. The  $\mathrm{YBr_2}$ and  $\mathrm{YCl_2}$ monolayers  have  been predicted to be stable FM semiconductors with GGA+$U$ ($U$=2.00 eV)\cite{p5}.  The AB-stacked $\mathrm{YBr_2}$/$\mathrm{YCl_2}$ heterojunction (see \autoref{f} (a)) is constructed with an AFM1 magnetic ground state.  The total magnetic moment of $\mathrm{YBr_2}$/$\mathrm{YCl_2}$ heterojunction is 0.00 $\mu_B$, which is further confirmed by  the fact that the IDOSs of  the two spin channels are the same within the band gap (see \autoref{f} (b)).
According to \autoref{f} (c) and (d),  the calculated band structures  show that  $\mathrm{YBr_2}$/$\mathrm{YCl_2}$ heterojunction possesses a pronounced spin-splitting and a type-II heterojunction (The type of the heterojunction depends on the magnitude of $U$.). Therefore, $\mathrm{YBr_2}$/$\mathrm{YCl_2}$ heterojunction can achieve fully-compensated ferrimagnetism.

We have also constructed AB-stacked $\mathrm{CrI_3}$/$\mathrm{CrBr_3}$ heterostructure (see FIG.S4\cite{bc})  using experimentally synthesized $\mathrm{CrI_3}$ and $\mathrm{CrBr_3}$ monolayers\cite{p2,p6}. Unfortunately, within a reasonable range of $U$, the $\mathrm{CrI_3}$/$\mathrm{CrBr_3}$ heterostructure always maintains a FM ground state. Nevertheless, we have also calculated its electronic structure with the AFM1 magnetic ordering. Its total magnetic moment is indeed 0.00 $\mu_B$, and there is a pronounced spin-splitting (see FIG.S4\cite{bc}) .
A large number of 2D FM monolayers have been theoretically predicted\cite{p7}, thus enabling the creation of a multitude of heterojunctions with zero-net-magnetization. The most crucial aspect here is that, to achieve fully-compensated ferrimagnetism, these heterojunctions should possess A-type AFM ordering, which can be simply estimated through an electron-counting rule\cite{p8}.

\textcolor[rgb]{0.00,0.00,1.00}{\textbf{Discussion and conclusion.---}}
The interface of vertical heterostructures is maintained by weak vdW forces, which do not require strict lattice matching and allow the combination of materials with different lattice constants. This is more conducive to realizing our proposal in a wider range of 2D FM materials.
Achieving altermagnetism through the stacking of bilayer has symmetry requirements\cite{k6,p9,p10}, that is to say, it is very sensitive to the stacking manner, which poses a challenge for experimental realization.  Our proposal to ahiceve  fully-compensated ferrimagnetism  does not have any symmetry requirements for the stacking manner. As long as the A-type AFM coupling is satisfied, a fully-compensated ferrimagnet with pronounced spin-splitting can be achieved. These advantages are highly conducive to the experimental realization of fully-compensated ferrimagnetism.

In summary, the  vertical heterostructures are proposed with two different but equally magnetized 2D ferromagnetic materials to achieve a fully-compensated ferrimagnet. This approach is insensitive to lattice matching and stacking manner, facilitating experimental realization. First-principles calculations show that $\mathrm{CrI_3}$/$\mathrm{CrGeTe_3}$ heterojunction can form a fully-compensated ferrimagnet with pronounced spin-splitting, and tensile strain can  enhance this state. Our work provides a feasible strategy for experimentally realizing fully-compensated ferrimagnetism.

\begin{acknowledgments}
This work is supported by Natural Science Basis Research Plan in Shaanxi Province of China  (2025JC-YBMS-008). Y.S.A. is supported by the Singapore Ministry of Education Academic Research Fund Tier 2 (Award No. MOE-T2EP50221-0019). We are grateful to Shanxi Supercomputing Center of China, and the calculations were performed on TianHe-2.
\end{acknowledgments}

\end{document}